\documentclass[twocolumn,showpacs,amsmath,amssymb,floatfix,superscriptaddress]{revtex4-1}
\usepackage{graphicx,times}
\begin{document}

\title{High-Field Superconductivity at an Electronic Topological Transition in URhGe}
\author{E. A. Yelland}
\affiliation{Scottish Universities Physics Alliance (SUPA), School of Physics and Astronomy and Centre for Science at Extreme Conditions, University of Edinburgh, Mayfield Road, Edinburgh EH9 3JZ, UK}
\affiliation{SUPA, School of Physics and Astronomy, University of St Andrews, North Haugh, St Andrews KY16 9SS, UK}
\author{J. M. Barraclough}
\affiliation{SUPA, School of Physics and Astronomy, University of St Andrews, North Haugh, St Andrews KY16 9SS, UK}
\author{W. Wang}
\affiliation{School of Engineering and Centre for Science at Extreme Conditions, University of Edinburgh, Mayfield Road, Edinburgh EH9 3JZ, UK}
\author{K. V. Kamenev}
\affiliation{School of Engineering and Centre for Science at Extreme Conditions, University of Edinburgh, Mayfield Road, Edinburgh EH9 3JZ, UK}
\author{A. D. Huxley}
\affiliation{Scottish Universities Physics Alliance (SUPA), School of Physics and Astronomy and Centre for Science at Extreme Conditions, University of Edinburgh, Mayfield Road, Edinburgh EH9 3JZ, UK}
\affiliation{SUPA, School of Physics and Astronomy, University of St Andrews, North Haugh, St Andrews KY16 9SS, UK}
\date{\today}

\begin{abstract}

The emergence of superconductivity at high magnetic fields in URhGe is regarded as a paradigm for new state formation approaching a quantum critical point. Until now, a divergence of the quasiparticle mass at the metamagnetic transition was considered essential for superconductivity to survive at magnetic fields above 30 tesla. Here we report the observation of quantum oscillations in URhGe revealing a tiny pocket of heavy quasiparticles that shrinks continuously with increasing magnetic field, and finally disappears at a topological Fermi surface transition close to or at the metamagnetic field. The quasiparticle mass decreases and remains finite, implying that the Fermi velocity vanishes due to the collapse of the Fermi wavevector. This offers a novel explanation for the re-emergence of superconductivity at extreme magnetic fields and makes URhGe the first proven example of a material where magnetic field-tuning of the Fermi surface, rather than quantum criticality alone, governs quantum phase formation.
\end{abstract}


\maketitle

The discovery and understanding of new quantum phases is a central theme of research in strongly correlated electron systems. Metals with narrow bandwidths including the $f$-electron heavy fermion (HF) materials show rich behaviour because their high density of states (DOS) promotes correlation effects and the energy scale for navigating the phase diagram is accessible with realistic magnetic fields and pressures. The HF metals UGe$_2$ \cite{Saxena00}, URhGe \cite{Aoki01} and UCoGe \cite{Huy07} attract particular interest because they show microscopic coexistence of superconductivity (SC) and ferromagnetism (FM), which are competing orders in conventional SC theories with opposite-spin pairs. They therefore offer the prospect of realizing the long-predicted metallic analogue of the A1 superfluid phase in $^3$He \cite{Fay80} in which magnetic fluctuations bind together quasiparticles with equal spin. Strong experimental evidence that the Cooper pairs are indeed equal-spin states in URhGe is provided by the sensitivity of SC to disorder and the magnitude and $T$ dependence of the critical field for destruction of SC \cite{Hardy05}.

\begin{figure}
\includegraphics*[clip=true,width=4.2in]{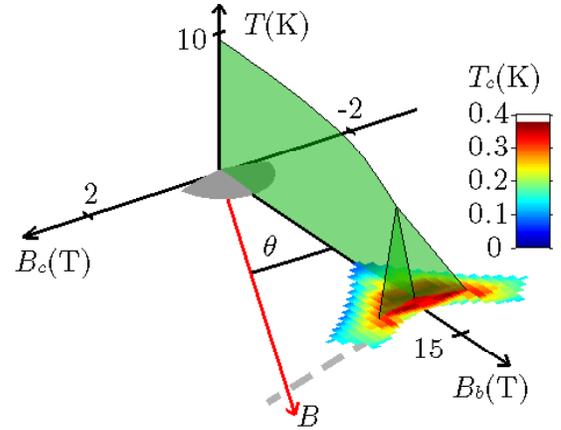}
\caption{Schematic phase diagram of URhGe for magnetic fields applied in the crystallographic $b,c$-plane. Below 9.5\,K URhGe is ferromagnetic with its moment parallel to the $c$-axis. An applied magnetic field causes the moment to rotate towards the field direction; changes are smooth except when crossing one of the sheets where a first order moment-rotation transition occurs. Two regions of superconductivity exist: one at low field and another at high field surrounding the point where the three transition surfaces meet. Our quantum oscillation results reveal a Fermi surface transition across the dashed line at $\theta=10^\circ$.}
\label{fig:URG_3d_phase_diag}
\end{figure}

The phase diagram of URhGe is shown schematically in Fig.~\ref{fig:URG_3d_phase_diag}. Ferromagnetism exists below $T=9.5$\,K with a spontaneous magnetic moment $M_\mathrm{c}=0.4\,\mu_\mathrm{B}$ parallel to the crystal $c$-axis. Bulk SC forms deep within the FM state \cite{Aoki01} at $T_\mathrm{c}=275$\,mK in the cleanest crystals. A magnetic field applied along $b$ first destroys SC but remarkably it reappears between 8\,T and $\approx\!12.5$\,T with a higher $T_\mathrm{c}$ than at zero field \cite{Levy05}. Measurements with $B$ tilted by an angle $\theta$ away from $b$ within the easy $b,c$-plane \cite{Levy05} suggest the surface of first order transitions separating $M_\mathrm{c}\,>\,0$ from $M_\mathrm{c}\,<\,0$ bifurcates at a tricritical point around 12\,T giving two surfaces $B_\mathrm{R}(\pm \theta,T)$ across which the moment rotates discontinuously towards $B$. These surfaces extend to angles a few degrees away from $b$, beyond which the transition is replaced by a cross-over. It has not been established if fluctuations associated with possible quantum critical points at the end of the first order lines provide the pairing interaction that drives SC. Independent of the microscopic origin of the superconductivity, the upper critical field of both SC pockets obeys a simple model for orbitally limited SC with (i) a $B$-independent anisotropy of the Fermi velocity and (ii) a $B$-dependent SC coherence length $\xi$ that continuously decreases as $B_\mathrm{R}$ is approached from the low or high field side \cite{Levy07}.

Here we report the direct observation of the Fermi surface (FS) in URhGe via the Shubnikov-de Haas (SdH) effect,  providing precise information on the FS geometry and quasiparticle mass in a crucial part of the phase diagram. The magnetic field dependence of the quantum oscillation (QO) frequency and amplitude, and the quasiparticle mass, together with a non-Fermi liquid form for the resistivity, suggest that one or more Fermi surface pockets vanish at a zero-temperature field-induced topological transition, also known as a Lifshitz transition (LT). A simple model consistent with our observations gives an orbital-limiting field that can explain the field-extent of the re-entrant SC in quantitative agreement with experiment.

\begin{figure}
\includegraphics*[clip=true,width=3.25in]{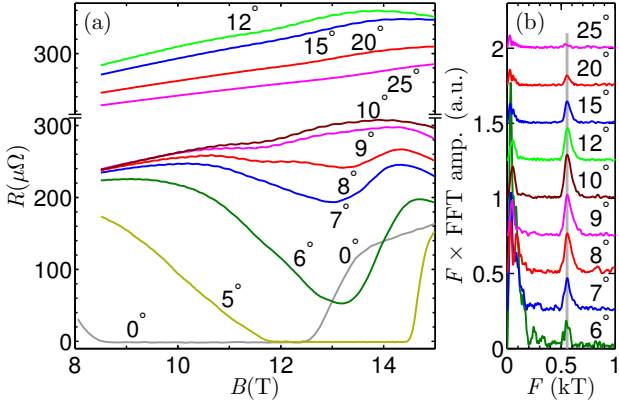}
\caption{Field-angle dependence of $R(B)$ at $T\approx 100$\,mK (a) Raw data for various angles $\theta$. The ripples visible at $\theta=10^\circ$ are Shubnikov-de Haas oscillations. (b) Fast Fourier transform of $R(B)$ after subtracting a linear background. The oscillations give a sharp peak marked by the vertical line at $F=555$\,T. This reaches a maximum amplitude at 10$^\circ$, where $R$ is also maximum, before disappearing at higher angles.}
\label{fig:URhGe_ang_dep}
\end{figure}

Our magnetoresistance measurements were made on a single crystal of URhGe with a residual resistance ratio $\rho(300\,\text{K})/\rho(T\,\rightarrow\,0\,\text{K})$=130 indicating a high degree of crystalline order (see Methods). Fig.\ \ref{fig:URhGe_ang_dep} shows the angle dependence of $R(B)$ as $B$ is rotated from $\hat{b}$ ($\theta=0^\circ$) towards $\hat{c}$ with $T\approx 100$\,mK. For $B\,\|\,\hat{b}$ a SC pocket exists in the range $8.5\leq B\leq 12.5$\,T \cite{Levy05}. On rotating $B$ towards $\hat{c}$, the $B$-width of the zero-resistance region shrinks continuously, reaching zero at $\theta\approx 5.5^\circ$. By $\theta=6^\circ$, a Fast Fourier transform (FFT) in $1/B$ of $R(B)$ reveals a sharp peak at $F=555$\,T that reaches a maximum amplitude at $\theta=10^\circ$. The periodicity and $T$ dependence of the oscillations are characteristic of quantum oscillations (QO) that occur at high magnetic fields in clean samples when the electronic states become confined to Landau tubes \cite{ShoenbergBook}.

QO in resistivity, also called Shubnikov-de Haas (SdH) oscillations, stem principally from the modulation of the electronic scattering rate \cite{ShoenbergBook}. A quantum mechanical treatment of electron scattering at high magnetic field for a 3-dimensional metal \cite{Adams59} leads to
\begin{eqnarray}
\tilde{\rho}/\rho & \approx & \alpha R_\mathrm{T} |\tilde{n}(\epsilon_\mathrm{F})|/n(\epsilon_\mathrm{F}), \nonumber \\
R_T & = &\frac{2 \pi^2 k_\mathrm{B} m^* T/e\hbar B}{\sinh (2 \pi^2 k_\mathrm{B} m^* T/e\hbar B)}
\label{eq:SdH_mag}
\end{eqnarray}
where $\tilde{\rho}$ and $\tilde{n}(\epsilon_\mathrm{F})$ are the oscillatory amplitudes of the resistivity $\rho$ and DOS $n(\epsilon_\mathrm{F})$ arising from the passage of Landau tubes through a particular extremal cross-section of the Fermi surface; $\alpha$ is a number $\sim\!1$ depending on the scattering mechanism and $R_T$ describes the attenuation of the oscillations due to thermal broadening of the Landau levels. $R_T$ is important because it allows the enhanced quasiparticle mass $m^*$ to be determined, providing an experimental probe of the same many-body interactions that enhance the specific heat. The fundamental component of $\tilde{n}(\epsilon_\mathrm{F})$ is given by the Lifshitz-Kosevich formula \cite{ShoenbergBook}
\begin{equation}
\tilde{n}(\epsilon_\mathrm{F}) \propto \frac{B^\frac{1}{2} R_\mathrm{D}}{\sqrt{\partial^2 \mathcal{A}/\partial k_\parallel^2}} \cos\left[\frac{2\pi F(B)}{B}+\phi \right],
\label{eq:SdH_DOSoscs}
\end{equation}
where $\mathcal{A}$ is the area of the orbit in $k$-space, $B$ is the magnetic induction \cite{Bcorr}, $R_\mathrm{D}$ is the Dingle factor describing damping due to scattering and $F(B)=(\hbar/2\pi e)\mathcal{A}(B)$ is the QO frequency. The curvature factor $\partial^2 \mathcal{A}/\partial k_\parallel^2$ accounts for the number of $k$-states that coherently contribute to the oscillatory amplitude. For ferromagnets such as URhGe opposite spins give rise to QOs with distinct frequencies so the spin interference factor that applies to paramagnetic metals is not present here.

The Fast Fourier transforms of our resistivity measurements for URhGe in Fig.~\ref{fig:URhGe_ang_dep}(b) show a clear peak at $F=555$\,T, well separated from spectral-weight below 200\,T (see Supplementary Information). This quantum oscillation component is the focus of our analysis. The frequency corresponds to a Fermi surface cross-section with a zero-field-projected area $\mathcal{A}=0.053$\,\AA$^{-2}$, $7\%$ of the Brillouin zone area. The absence of any resolvable angle dependence of $F$ in Fig.~\ref{fig:URhGe_ang_dep}(b) provides an upper limit $|F(20^\circ)-F(0^\circ)|\leq 10$\,T that excludes a locally 2D FS region or a FS neck. The data are consistent with a spherical or ellipsoidal pocket that is nearly circular in the $b,c$-plane and in the first case the pocket's small size means that it would contain only $\pi k_\mathrm{F}^3/(3 a^*b^*c^*)=2.1\times 10^{-3}$ carriers per U.

\begin{figure}
\includegraphics[clip=true,scale=0.394]{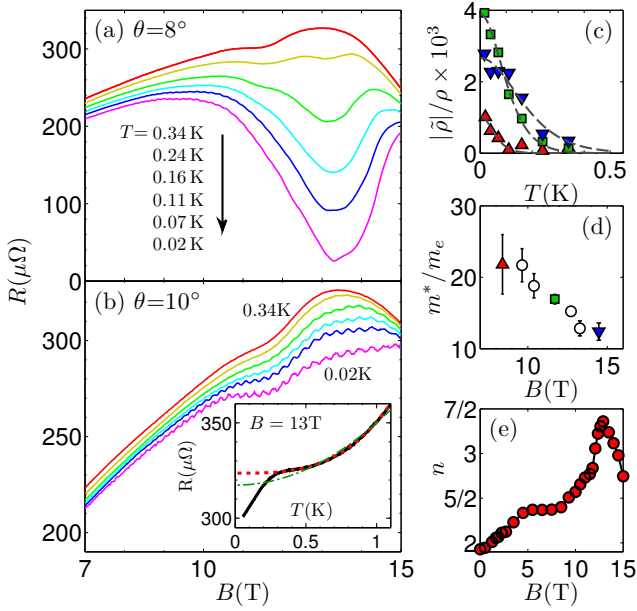}
\caption{Temperature dependence of the quantum oscillations. (a,b): $R(B)$ at $\theta\!=\!8^\circ$ and $10^\circ$ for various $T$. Inset: $R(T)$ at $10^\circ$ and $13\,$T (solid curve) with fits described in the text. (c) QO amplitude versus $T$ for the 10$^\circ$ data at selected values of $B$ (symbols) and fits to the Lifshitz-Kosevich form (dashed lines). (d) The $B$-dependence of the quasiparticle mass $m^*$ from the fits in (c). (e) Exponent $n$ from fits of $R(T)$ to $\rho=\rho_0+A T^n$ for $0.4\!\leq\! T\!\leq\!1.1$\,K at $\theta=10^\circ$.}
\label{fig:URG_QO_Tdep}
\end{figure}

The temperature dependence of $R(B)$ is shown in Fig.~\ref{fig:URG_QO_Tdep}(a,b) at $\theta=8^\circ$ and $\theta=10^\circ$ where the QO signal is strongest. We have extracted the amplitude envelope of the oscillating part of $R(B)$ as a function of $B$ and $T$ (see Supplementary Information). The $T$-dependent amplitude at fixed $B$ was then fitted to $R_T$ in Eq.~\ref{eq:SdH_mag} as shown for representative values of $B$ in Fig.~\ref{fig:URG_QO_Tdep}. The fits provide the quasiparticle mass $m^*(B)$ shown in Fig.~\ref{fig:URG_QO_Tdep}(d) which decreases from $\approx 22m_e$ at 8\,T to $\approx 12m_e$ at 15\,T. For a spherical pocket, the contribution to the low temperature linear coefficient in the specific heat, $\gamma$, is $\gamma=n_\mathrm{s} (k_\mathrm{B}^2/6\hbar^2) k_\mathrm{F}m^*$ per unit volume where $n_\mathrm{s}$ is the number of copies of the FS sheet in the BZ and there is no spin degeneracy. For $m^*=20m_e$ and $\mathcal{A}=0.053$\,\AA$^{-2}$, $\gamma_\mathrm{QO}=2.3$\,mJ\,mol$^{-1}$\,K$^{-2}$ per sheet. There are no measurements of the heat capacity at this field angle for comparison, but magnetization \cite{Aoki10_2} and a.c.\ calorimetry measurements \cite{Levy09} suggest that the total heat capacity coefficient at 9\,T and $\theta\,\approx\,0^\circ$ is little different from the zero-field value $\gamma_\mathrm{TOT}=160$\,mJ\,mol$^{-1}$\,K$^{-2}$ \cite{Aoki01}. Since the contribution from the pocket(s) we observe is a lot smaller than this, other Fermi surfaces not detected in our study must be present.

In the standard Fermi liquid description of the metallic state, the linear term in the specific heat is closely connected to the $A$ coefficient in the low temperature electrical resistivity $\rho=\rho_0+A T^2$. Generally, $\gamma \propto m^*$ while $A\propto m^{*2}$. In our measurements on URhGe at $\theta = 10^\circ$, we find that the $T$-dependence of the non-oscillatory background resistivity does indeed weaken with field as $m^*$ decreases, but it does not have the usual Fermi liquid form and therefore cannot be simply related to changes of the electronic heat capacity. It was reported \cite{Aoki10} in UCoGe that $m^*$ also decreases with increasing field at the same time the $T$ dependence of the resistivity weakens, but in that case the resistivity did follow a Fermi liquid form. For $B\rightarrow 0$\,T in URhGe the usual Fermi liquid form is recovered above $T_\mathrm{c}$ but strong deviations occur at high field where distinct low ($T\!<\!T^*$) and high $(T\!>\!T^*)$ temperature regions exist that have nearly linear and super-linear $T$ dependences as in Fig.\ \ref{fig:URG_QO_Tdep}(b) (inset) at 13\,T. The crossover temperature $T^*\,\lesssim\,0.4\,$K for all fields. If a $T^2$ temperature dependence is imposed to fit the data as illustrated by the dash-dotted curve in the inset of Fig.\ \ref{fig:URG_QO_Tdep}(b), the experimental resistivity has an initial upward deviation from this with decreasing temperature as well as a subsequent sharp downturn below $T^*$. The upturn is present at all fields and therefore cannot be attributed to low frequency QO's which would give contributions to the resistivity that oscillate from positive to negative as a function of field; a simple $T^2$ dependence above $T^*$ in combination with QO's cannot therefore explain our data. In contrast, a fit of the form $\rho-\rho_0\propto T^n$ with $n$ free to vary describes the data well in the higher $T$ region ($0.4\,\leq\,T\,\leq\,1.1$\,K). In this case only the downturn below $T^*$ requires additional explanation. The continuous evolution with angle of the slow undulations in Fig.~ \ref{fig:URhGe_ang_dep} to a state that is almost resistanceless at 20\,mK and $\theta = 8^\circ$ [Fig.~\ref{fig:URG_QO_Tdep}(a)], and is resistanceless for smaller $\theta$, hints that the behaviour below $T^*$ relates strongly to the proximity to SC, for example, a partial transition into a superconducting state. Applying the floating-power fit yields the exponents shown in Fig.\ \ref{fig:URG_QO_Tdep}(e): $n=2.0$ at zero field, but $n$ increases to $\approx 2.4$ at 8\,T before jumping sharply to $\approx 3.4$ at 13\,T and then decreasing again beyond the magnetic crossover field \cite{Levy05}. A non-$T^2$ power law with $n=5/2$ or $7/2$ is consistent with proximity to a field-induced 3D Lifshitz transition \cite{Blanter94,Katsnelson00,Hackl11}.

\begin{figure}
\includegraphics[clip=true,scale=0.394]{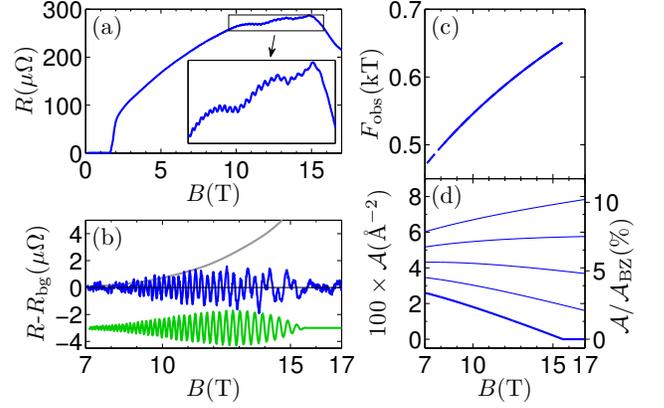}
\caption{Magnetic field dependence of the quantum oscillations. (a) $R(B)$ at $T=20$\,mK for $B\leq 17$\,T and $\theta \approx 10^\circ$. The oscillations disappear suddenly at $B\approx 15.5$\,T where the non-oscillatory part of $R$ also drops sharply. (b) Residual after subtracting a smoothly varying background (blue curve) and Lifshitz-Kosevich model calculations for a Fermi surface orbit that shrinks to zero size (green curve) and the amplitude envelope expected for a constant orbit size (grey curve). (c) Observed quantum oscillation frequency $F_\mathrm{obs}(B)$. (d) Forms of the $B$ dependent Fermi surface area that are consistent with $F_\mathrm{obs}(B)$. The bold line shows $\mathcal{A}(B)$ vanishing at 15.5\,T corresponding to a Lifshitz transition where a Fermi pocket shrinks to a point or a neck pinches off.}
\label{fig:URG_QO_Bdep}
\end{figure}

The magnetic field dependence of $R$ at $T=20$\,mK is shown in Fig.~\ref{fig:URG_QO_Bdep} up to $17$\,T. The QO component centered on $555$\,T has been isolated from the $B$ dependent background by subtraction of a smooth function [residuals shown in Fig.~\ref{fig:URG_QO_Bdep}(b)]. There are $>\!30$ consecutive oscillations resolvable in the range $8.2\leq B \leq 15.5$\,T, allowing a precise comparison with the LK theory expressions Eqs.~\ref{eq:SdH_mag} and \ref{eq:SdH_DOSoscs}. An LK model calculation (details in Supplementary Information) is shown in Fig.~\ref{fig:URG_QO_Bdep}(b). The oscillating part of the model expression contains 3 adjustable parameters controlling the oscillation phase, frequency and the $B$-dependence of the frequency. Crucially, the accurate phase relation between fit and data over the entire range of $B$ can only be achieved by including a $B$-dependent term in the frequency. In a paramagnet $F(B)$ in Eq.\,\ref{eq:SdH_DOSoscs} depends linearly on B and can be replaced by $F(0)$, with the $B$-linear part absorbed in the QO phase. However $B$-dependent QO frequencies are expected in any itinerant magnet in which $M$ varies nonlinearly with $B$ and have been observed in a number of $f$- and $d$-electron materials, both FM and non-FM, e.g. UPt$_3$ \cite{Julian92}, ZrZn$_2$ \cite{vanRuitenbeek82}, Sr$_3$Ru$_2$O$_7$ \cite{Mercure10}, and YbRh$_2$Si$_2$ \cite{Rourke08}. Our observed QO frequency $F_\mathrm{obs}$ is shown in Fig.~\ref{fig:URG_QO_Bdep}(c). The actual FS cross-sectional area $\mathcal{A}(B)$ is related to $F_\mathrm{obs}$ by $F_\mathrm{obs}(B) \propto \mathcal{A}(B)-B\mathrm{d}\mathcal{A}/\mathrm{d}B$ \cite{Julian92}. Fig.~\ref{fig:URG_QO_Bdep}(d) shows forms of $\mathcal{A}(B)$ that are consistent with $F_\mathrm{obs}$, differing from each other only in the value of the undetermined $B$-linear term. The lowest curve for $\mathcal{A}(B)$ shows it reaching zero at $15.5$\,T, where we suggest a magnetic field induced Lifshitz transition occurs; this provides a natural explanation for the sudden loss of oscillatory signal at high field.

The dependence of the QO amplitude on $B$ contains information about quasiparticle scattering via a mean free path $\ell_\mathrm{Q}$ that enters the Dingle factor $R_\mathrm{D}=\exp(-\pi\sqrt{2\hbar F/e}/B\ell_\mathrm{Q})$ in Eq.~\ref{eq:SdH_DOSoscs}. Comparing the amplitude envelope of the oscillatory data in Fig.~\ref{fig:URG_QO_Bdep}(b) to the LK model with $\ell_\mathrm{Q}=550$\,\AA\ and $\mathcal{A}=0.043$\,\AA$^{-2}$, independent of $B$, we see that the model tracks the data up to $\sim\!11$\,T but at higher $B$ the oscillations rapidly become weaker than predicted and abruptly vanish within experimental resolution at $15.5$\,T. Note that the effect of the decrease in the background resistance above 15\,T is included in the calculation. The lower curve shows the LK result for the case of a Lifshitz transition where $\mathcal{A}(B)$ is given by the lowest curve in Fig.~\ref{fig:URG_QO_Bdep}(d). Using a $B$-independent mean free path $\ell_\mathrm{Q}=550\pm100$\,\AA\, the calculated QO amplitude is in good agreement with experiment over the entire range of $B$  (see Supplementary Information for further details of the calculation). Physically the loss of amplitude occurs both because carriers are slowing down approaching the LT and because they are diminishing in number. Without attempting to capture the details of multi-band magnetotransport, this simple argument shows that a $B$-induced decrease of the Fermi wavevector culminating in the disappearance of the pocket at a LT can explain the otherwise anomalous $B$ dependent QO amplitude. Other potential explanations can be eliminated as described in Supplementary Information.

\begin{figure}
\includegraphics[clip=true,scale=0.394]{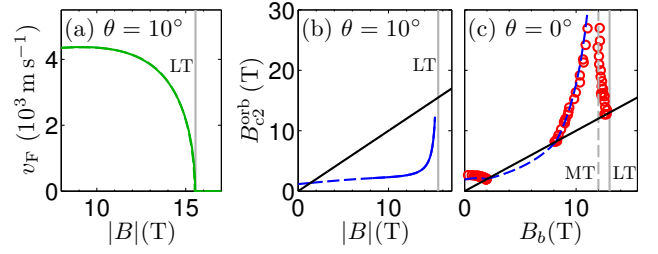}
\caption{Magnetic field dependence of the Fermi velocity $v_\mathrm{F}$ and orbital-limiting field $B_\mathrm{c2}^\mathrm{orb}$ for the Fermi pocket detected in this study. (a) $v_\mathrm{F}(B)=\hbar k_\mathrm{F}/m^{*}$ with $k_\mathrm{F}$ and $m^*$ determined by quantum oscillations at $\theta=10^\circ$. $v_\mathrm{F}\rightarrow 0$ at the Lifshitz transition (LT) where $k_\mathrm{F}\rightarrow 0$. (b) $B_\mathrm{c2}^\mathrm{orb}=\Phi_{0}/2\pi\xi^{2}$ at $\theta=10^\circ$ calculated from $v_\mathrm{F}$ (solid curve; dashed curve uses extrapolated values of $k_\mathrm{F}$ and $m^*$). High-$B$ superconductivity cannot occur because $B_\mathrm{c2}^\mathrm{orb}\!<\!|B|$. (c) Calculated $B_\mathrm{c2}^\mathrm{orb}$ at $\theta=0^\circ$ if a LT occurs at 13\,T for a $B$ independent mass $m^*=40\,m_e$  (dashed line). MT marks the magnetic transition. Now $B_\mathrm{c2}^\mathrm{orb}\!>\!B_b$ at high $B$ so orbital limiting does not destroy superconductivity for $8\!\lesssim\!B\!\lesssim\!13$\,T. Symbols show measured values of the superconducting $B_\mathrm{c2}$ \cite{Levy07} which are remarkably well described by the model.}

\label{fig:URG_Hc2_orb}
\end{figure}

To examine whether our data are consistent with a Zeeman-driven Lifshitz transition we consider for definiteness a vanishing pocket of minority spin electrons but the discussion also applies to majority spin holes. As $B$ is increased, the Fermi level relative to the bottom of the pocket decreases, and its $B$-derivative ${\partial(\epsilon_\mathrm{F}-\epsilon_0)}/{\partial B} = 2 \mu_\text{eff}$ can be used to estimate $\mu_\text{eff}$, the component of the effective electronic moment parallel to the field. The factor of 2 applies when the Fermi level is effectively pinned by a dominant DOS on the majority spin band. We consider two possibilities: (i) the density of states of the pocket-band is unaffected as $\epsilon_\mathrm{F}-\epsilon_0$ changes with field, giving $2 \mu_\text{eff} = (1/m^*)\partial \mathcal{A}/\partial{B}$; (ii) the bandwidth renormalization changes with field but the band maintains a parabolic dispersion for which $2 \mu_\text{eff} = \partial (\mathcal{A}/m^*)/\partial{B}$. The true case might lie between these two extremes. The QO results provide both $\mathcal{A}(B)$ and an enhanced quasiparticle mass $m^*(B)=(\hbar^2/2\pi) \mathrm{d}\mathcal{A}/\mathrm{d}E$.  For both cases we find $\mu_\mathrm{eff}$ is small at low field where the component of the magnetic moment $\|B$ is known to be small, but it increases to $\approx 3\,\mu_\mathrm{B}$ at 15T where the moment has rotated towards the field direction. This value is approximately a factor of 2 larger than expected for U moments \cite{Freeman76}, but would be in good agreement for a Wilson ratio of 2 appropriate for the Kondo effect in which the susceptibility enhancement is a factor 2 larger than the DOS enhancement. The relationship between the field- and energy-scale of the unusual features in our QO data thus strongly supports the interpretation in terms of a Lifshitz transition.

We now discuss the consequences of our findings for field-induced superconductivity. In an equal-spin-paired superconductor the upper critical field is expected to be orbitally limited to a value $B_\mathrm{c2}^\mathrm{orb}=\Phi_0/(2\pi\xi^2)$ which is the maximum field that a type II superconductor can sustain in its mixed state before the normal cores of the flux vortices occupy the entire volume and superconductivity is destroyed. In BCS theory $\xi=\hbar v_\mathrm{F}/\pi\Delta$ where $\Delta$ measures the excitation gap in the SC state, so at sufficiently high magnetic fields in a multi-band metal the superconducting coherence length $\xi$ would be associated with the band with lowest Fermi velocity. Fig.~\ref{fig:URG_Hc2_orb}(a) shows the quasiparticle velocity $v_\mathrm{F}=h k_\mathrm{F}/m^*$ on the shrinking orbit calculated from the experimental values of $m^*$ and $k_\mathrm{F}$. The values are low and decrease from $4.4\times 10^3$\,m\,s$^{-1}$ at 9\,T to zero at the LT. To estimate $\xi$ in absolute units we take $\Delta=2\,k_\mathrm{B}T_\mathrm{c}$ with $T_\mathrm{c}(B)$ chosen to interpolate linearly between the measured zero-field value and the observed maximum value of 0.45\,K at 12\,T for $B\|b$. The resulting orbital limiting field diverges on approaching the Lifshitz transition as shown in Fig.~\ref{fig:URG_Hc2_orb}(b). For $\theta=10^\circ$ where we know $v_\mathrm{F}$ from the QOs, the orbital-limiting field just fails to exceed the applied field up to a cut-off at $k_\mathrm{F}\xi=1$ beyond which the BCS expression for $\xi$ cannot be applied. This is consistent with the observed absence of SC at high $B$ at $10^\circ$. To predict what could occur at $\theta=0^\circ$, we force $\mathcal{A}(B)$ to vanish at 13\,T where the high-$B$ SC is destroyed. The resulting $\xi(B)$ leads to a region at high field in Fig.~\ref{fig:URG_Hc2_orb}(c) where $B_\mathrm{c2}^\mathrm{orb}$ exceeds $B$ and equal-spin-paired SC is not prevented by orbital limiting. This calculation uses a $B$-independent mass $m^*=40 m_e$ (further details are in Supplementary Information). The absolute magnitude and $B$ dependence are remarkably close to experimental values of the superconducting $B_\mathrm{c2}$ \cite{Levy07} suggesting that the slowing down of quasiparticles at a Lifshitz transition would allow superconductivity to survive above 30\,T, without a divergence of the quasiparticle mass.

Recent experiments on CeRu$_2$Si$_2$ \cite{Daou06} suggest that the metamagnetic transition in this material is also associated with a sheet of the Fermi surface shrinking continuously to zero size and the same physics may underly the field-induced transition to antiferromagnetism in YbRh$_2$Si$_2$ \cite{Hackl11,Kusminskiy08}. In both these cases the LT separates two phases (both Fermi liquids in the low $T$ limit) but new state formation around the transition has not been observed. Our results on URhGe provide the first example where the presence of a LT enables phase formation in the vicinity of a QCP, in this case by creating conditions favorable for high-field superconductivity. It remains an open question whether the Liftshitz transition also plays a role in shaping the spectrum of magnetic fluctuations responsible for superconducting pairing in URhGe. This highlights the need for a theory of superconductivity in URhGe that includes the effects of a Lifshitz transition alongside quantum criticality, addressing both the changes to the spectrum of magnetic fluctuations \cite{Yamaji07} and the existence of critically slow fermionic quasiparticles. More generally, topological transitions of the Fermi surface may be commoner than presently thought in narrow-band metals and may offer a route to quantum phase formation that deserves more attention.

\textbf{Methods.} Magnetoresistance measurements were made using two different setups. The first used piezoelectric rotators to control the field angle with relative precision $\sim 0.02^\circ$ about 2 axes for $T\gtrsim 80$\,mK and $B\leq 17$\,T with a field-calibrated RuO$_2$ resistance thermometer mounted next to the sample. In the second setup the sample was aligned by Laue diffraction and rigidly mounted to a stage that was strongly thermally coupled to a zero-field thermometer. This provided a cross-check on the in-field thermometry of the first set-up and allowed temperatures down to 20\,mK. A standard a.c.\! lock-in technique and a low temperature transformer were used giving sensitivity to QO signals of amplitude $\geq 60\,$n$\Omega\cong 3$\,pV; the measurement current was always $\leq 100$\,$\mu$A.

%bbl here

\textbf{Acknowledgements.} The sample studied was derived from material kindly supplied by the CEA-Grenoble. Research support was provided by the Engineering and Physical Sciences Research Council and the Royal Society.

\textbf{Author contributions.} EAY set up and made the measurements, analysed the data and wrote the manuscript with input from ADH. JMB contributed to the data acquisition and analysis. WW and KVK designed the 2-axis rotator stage.

%%%%
%merlin.mbs apsrev4-1.bst 2010-07-25 4.21a (PWD, AO, DPC) hacked
%Control: key (0)
%Control: author (72) initials jnrlst
%Control: editor formatted (1) identically to author
%Control: production of article title (-1) disabled
%Control: page (0) single
%Control: year (1) truncated
%Control: production of eprint (0) enabled
%

%%%%


\begin{thebibliography}{24}%
\makeatletter
\providecommand \@ifxundefined [1]{%
 \@ifx{#1\undefined}
}%
\providecommand \@ifnum [1]{%
 \ifnum #1\expandafter \@firstoftwo
 \else \expandafter \@secondoftwo
 \fi
}%
\providecommand \@ifx [1]{%
 \ifx #1\expandafter \@firstoftwo
 \else \expandafter \@secondoftwo
 \fi
}%
\providecommand \natexlab [1]{#1}%
\providecommand \enquote  [1]{``#1''}%
\providecommand \bibnamefont  [1]{#1}%
\providecommand \bibfnamefont [1]{#1}%
\providecommand \citenamefont [1]{#1}%
\providecommand \href@noop [0]{\@secondoftwo}%
\providecommand \href [0]{\begingroup \@sanitize@url \@href}%
\providecommand \@href[1]{\@@startlink{#1}\@@href}%
\providecommand \@@href[1]{\endgroup#1\@@endlink}%
\providecommand \@sanitize@url [0]{\catcode `\\12\catcode `\$12\catcode
  `\&12\catcode `\#12\catcode `\^12\catcode `\_12\catcode `\%12\relax}%
\providecommand \@@startlink[1]{}%
\providecommand \@@endlink[0]{}%
\providecommand \url  [0]{\begingroup\@sanitize@url \@url }%
\providecommand \@url [1]{\endgroup\@href {#1}{\urlprefix }}%
\providecommand \urlprefix  [0]{URL }%
\providecommand \Eprint [0]{\href }%
\providecommand \doibase [0]{http://dx.doi.org/}%
\providecommand \selectlanguage [0]{\@gobble}%
\providecommand \bibinfo  [0]{\@secondoftwo}%
\providecommand \bibfield  [0]{\@secondoftwo}%
\providecommand \translation [1]{[#1]}%
\providecommand \BibitemOpen [0]{}%
\providecommand \bibitemStop [0]{}%
\providecommand \bibitemNoStop [0]{.\EOS\space}%
\providecommand \EOS [0]{\spacefactor3000\relax}%
\providecommand \BibitemShut  [1]{\csname bibitem#1\endcsname}%
\let\auto@bib@innerbib\@empty
%</preamble>
\bibitem [{\citenamefont {Saxena}\ \emph {et~al.}(2000)\citenamefont {Saxena},
  \citenamefont {Agarwal}, \citenamefont {Ahilan}, \citenamefont {Grosche},
  \citenamefont {Haselwimmer}, \citenamefont {Steiner}, \citenamefont {Pugh},
  \citenamefont {Walker}, \citenamefont {Julian}, \citenamefont {Monthoux},
  \citenamefont {Lonzarich}, \citenamefont {Huxley}, \citenamefont {Sheikin},
  \citenamefont {Braithwaite},\ and\ \citenamefont {Flouquet}}]{Saxena00}%
  \BibitemOpen
  \bibfield  {author} {\bibinfo {author} {\bibfnamefont {S.~S.}\ \bibnamefont
  {Saxena}}, \bibinfo {author} {\bibfnamefont {P.}~\bibnamefont {Agarwal}},
  \bibinfo {author} {\bibfnamefont {K.}~\bibnamefont {Ahilan}}, \bibinfo
  {author} {\bibfnamefont {F.~M.}\ \bibnamefont {Grosche}}, \bibinfo {author}
  {\bibfnamefont {R.~K.~W.}\ \bibnamefont {Haselwimmer}}, \bibinfo {author}
  {\bibfnamefont {M.~J.}\ \bibnamefont {Steiner}}, \bibinfo {author}
  {\bibfnamefont {E.}~\bibnamefont {Pugh}}, \bibinfo {author} {\bibfnamefont
  {I.~R.}\ \bibnamefont {Walker}}, \bibinfo {author} {\bibfnamefont {S.~R.}\
  \bibnamefont {Julian}}, \bibinfo {author} {\bibfnamefont {P.}~\bibnamefont
  {Monthoux}}, \bibinfo {author} {\bibfnamefont {G.~G.}\ \bibnamefont
  {Lonzarich}}, \bibinfo {author} {\bibfnamefont {A.}~\bibnamefont {Huxley}},
  \bibinfo {author} {\bibfnamefont {I.}~\bibnamefont {Sheikin}}, \bibinfo
  {author} {\bibfnamefont {D.}~\bibnamefont {Braithwaite}}, \ and\ \bibinfo
  {author} {\bibfnamefont {J.}~\bibnamefont {Flouquet}},\ }\href@noop {}
  {\bibfield  {journal} {\bibinfo  {journal} {Nature}\ }\textbf {\bibinfo
  {volume} {406}},\ \bibinfo {pages} {587} (\bibinfo {year}
  {2000})}\BibitemShut {NoStop}%
\bibitem [{\citenamefont {Aoki}\ \emph {et~al.}(2001)\citenamefont {Aoki},
  \citenamefont {Huxley}, \citenamefont {Ressouche}, \citenamefont
  {Braithwaite}, \citenamefont {Flouquet}, \citenamefont {Brison},
  \citenamefont {Lhotel},\ and\ \citenamefont {Paulsen}}]{Aoki01}%
  \BibitemOpen
  \bibfield  {author} {\bibinfo {author} {\bibfnamefont {D.}~\bibnamefont
  {Aoki}}, \bibinfo {author} {\bibfnamefont {A.}~\bibnamefont {Huxley}},
  \bibinfo {author} {\bibfnamefont {E.}~\bibnamefont {Ressouche}}, \bibinfo
  {author} {\bibfnamefont {D.}~\bibnamefont {Braithwaite}}, \bibinfo {author}
  {\bibfnamefont {J.}~\bibnamefont {Flouquet}}, \bibinfo {author}
  {\bibfnamefont {J.-P.}\ \bibnamefont {Brison}}, \bibinfo {author}
  {\bibfnamefont {E.}~\bibnamefont {Lhotel}}, \ and\ \bibinfo {author}
  {\bibfnamefont {C.}~\bibnamefont {Paulsen}},\ }\href@noop {} {\bibfield
  {journal} {\bibinfo  {journal} {Nature}\ }\textbf {\bibinfo {volume} {413}},\
  \bibinfo {pages} {613} (\bibinfo {year} {2001})}\BibitemShut {NoStop}%
\bibitem [{\citenamefont {Huy}\ \emph {et~al.}(2007)\citenamefont {Huy},
  \citenamefont {Gasparini}, \citenamefont {de~Nijs}, \citenamefont {Huang},
  \citenamefont {Klaase}, \citenamefont {Gortenmulder}, \citenamefont
  {de~Visser}, \citenamefont {Hamann}, \citenamefont {G\"{o}rlach},\ and\
  \citenamefont {v.~L\"{o}hneysen}}]{Huy07}%
  \BibitemOpen
  \bibfield  {author} {\bibinfo {author} {\bibfnamefont {N.~T.}\ \bibnamefont
  {Huy}}, \bibinfo {author} {\bibfnamefont {A.}~\bibnamefont {Gasparini}},
  \bibinfo {author} {\bibfnamefont {D.~E.}\ \bibnamefont {de~Nijs}}, \bibinfo
  {author} {\bibfnamefont {Y.}~\bibnamefont {Huang}}, \bibinfo {author}
  {\bibfnamefont {J.~C.~P.}\ \bibnamefont {Klaase}}, \bibinfo {author}
  {\bibfnamefont {T.}~\bibnamefont {Gortenmulder}}, \bibinfo {author}
  {\bibfnamefont {A.}~\bibnamefont {de~Visser}}, \bibinfo {author}
  {\bibfnamefont {A.}~\bibnamefont {Hamann}}, \bibinfo {author} {\bibfnamefont
  {T.}~\bibnamefont {G\"{o}rlach}}, \ and\ \bibinfo {author} {\bibfnamefont
  {H.}~\bibnamefont {v.~L\"{o}hneysen}},\ }\href@noop {} {\bibfield  {journal}
  {\bibinfo  {journal} {Phys. Rev. Lett.}\ }\textbf {\bibinfo {volume} {99}},\
  \bibinfo {pages} {067006} (\bibinfo {year} {2007})}\BibitemShut {NoStop}%
\bibitem [{\citenamefont {Fay}\ and\ \citenamefont {Appel}(1980)}]{Fay80}%
  \BibitemOpen
  \bibfield  {author} {\bibinfo {author} {\bibfnamefont {D.}~\bibnamefont
  {Fay}}\ and\ \bibinfo {author} {\bibfnamefont {J.}~\bibnamefont {Appel}},\
  }\href {\doibase 10.1103/PhysRevB.22.3173} {\bibfield  {journal} {\bibinfo
  {journal} {Phys. Rev. B}\ }\textbf {\bibinfo {volume} {22}},\ \bibinfo
  {pages} {3173} (\bibinfo {year} {1980})}\BibitemShut {NoStop}%
\bibitem [{\citenamefont {Hardy}\ and\ \citenamefont {Huxley}(2005)}]{Hardy05}%
  \BibitemOpen
  \bibfield  {author} {\bibinfo {author} {\bibfnamefont {F.}~\bibnamefont
  {Hardy}}\ and\ \bibinfo {author} {\bibfnamefont {A.~D.}\ \bibnamefont
  {Huxley}},\ }\href@noop {} {\bibfield  {journal} {\bibinfo  {journal} {Phys.
  Rev. Lett.}\ }\textbf {\bibinfo {volume} {94}},\ \bibinfo {pages} {247006}
  (\bibinfo {year} {2005})}\BibitemShut {NoStop}%
\bibitem [{\citenamefont {L\'{e}vy}\ \emph {et~al.}(2005)\citenamefont
  {L\'{e}vy}, \citenamefont {Sheikin}, \citenamefont {Grenier},\ and\
  \citenamefont {Huxley}}]{Levy05}%
  \BibitemOpen
  \bibfield  {author} {\bibinfo {author} {\bibfnamefont {F.}~\bibnamefont
  {L\'{e}vy}}, \bibinfo {author} {\bibfnamefont {I.}~\bibnamefont {Sheikin}},
  \bibinfo {author} {\bibfnamefont {B.}~\bibnamefont {Grenier}}, \ and\
  \bibinfo {author} {\bibfnamefont {A.~D.}\ \bibnamefont {Huxley}},\
  }\href@noop {} {\bibfield  {journal} {\bibinfo  {journal} {Science}\ }\textbf
  {\bibinfo {volume} {309}},\ \bibinfo {pages} {1343} (\bibinfo {year}
  {2005})}\BibitemShut {NoStop}%
\bibitem [{\citenamefont {L\'{e}vy}\ \emph {et~al.}(2007)\citenamefont
  {L\'{e}vy}, \citenamefont {Sheikin},\ and\ \citenamefont {Huxley}}]{Levy07}%
  \BibitemOpen
  \bibfield  {author} {\bibinfo {author} {\bibfnamefont {F.}~\bibnamefont
  {L\'{e}vy}}, \bibinfo {author} {\bibfnamefont {I.}~\bibnamefont {Sheikin}}, \
  and\ \bibinfo {author} {\bibfnamefont {A.}~\bibnamefont {Huxley}},\
  }\href@noop {} {\bibfield  {journal} {\bibinfo  {journal} {Nature Physics}\
  }\textbf {\bibinfo {volume} {3}},\ \bibinfo {pages} {460} (\bibinfo {year}
  {2007})}\BibitemShut {NoStop}%
\bibitem [{\citenamefont {Shoenberg}(1984)}]{ShoenbergBook}%
  \BibitemOpen
  \bibfield  {author} {\bibinfo {author} {\bibfnamefont {D.}~\bibnamefont
  {Shoenberg}},\ }\href@noop {} {\emph {\bibinfo {title} {Magnetic Oscillations
  in Metals}}}\ (\bibinfo  {publisher} {Cambridge University Press},\ \bibinfo
  {year} {1984})\BibitemShut {NoStop}%
\bibitem [{\citenamefont {Adams}\ and\ \citenamefont
  {Holstein}(1959)}]{Adams59}%
  \BibitemOpen
  \bibfield  {author} {\bibinfo {author} {\bibfnamefont {E.~N.}\ \bibnamefont
  {Adams}}\ and\ \bibinfo {author} {\bibfnamefont {T.~D.}\ \bibnamefont
  {Holstein}},\ }\href@noop {} {\bibfield  {journal} {\bibinfo  {journal} {J.
  Phys. Chem. Solids}\ }\textbf {\bibinfo {volume} {10}},\ \bibinfo {pages}
  {254} (\bibinfo {year} {1959})}\BibitemShut {NoStop}%
\bibitem [{Bco()}]{Bcorr}%
  \BibitemOpen
  \href@noop {} {}\bibinfo {note} {We have estimated the difference between the
  applied field $B_\mathrm{app}$ and internal field $B_\mathrm{int}$ using
  $M(B)$ of URhGe measured at angles close to those reported here
  \cite{Levy05}. Resulting shifts of QO frequency are always $<15$\,T and
  therefore negligible.}\BibitemShut {Stop}%
\bibitem [{\citenamefont {Aoki}\ \emph {et~al.}(2010)\citenamefont {Aoki},
  \citenamefont {Matsuda}, \citenamefont {Hardy}, \citenamefont {Meingast},
  \citenamefont {Taufour}, \citenamefont {Hassinger}, \citenamefont {Sheikin},
  \citenamefont {Paulsen}, \citenamefont {Knebel}, \citenamefont {Kotegawa},\
  and\ \citenamefont {Flouquet}}]{Aoki10_2}%
  \BibitemOpen
  \bibfield  {author} {\bibinfo {author} {\bibfnamefont {D.}~\bibnamefont
  {Aoki}}, \bibinfo {author} {\bibfnamefont {T.~D.}\ \bibnamefont {Matsuda}},
  \bibinfo {author} {\bibfnamefont {F.}~\bibnamefont {Hardy}}, \bibinfo
  {author} {\bibfnamefont {C.}~\bibnamefont {Meingast}}, \bibinfo {author}
  {\bibfnamefont {V.}~\bibnamefont {Taufour}}, \bibinfo {author} {\bibfnamefont
  {E.}~\bibnamefont {Hassinger}}, \bibinfo {author} {\bibfnamefont
  {I.}~\bibnamefont {Sheikin}}, \bibinfo {author} {\bibfnamefont
  {C.}~\bibnamefont {Paulsen}}, \bibinfo {author} {\bibfnamefont
  {G.}~\bibnamefont {Knebel}}, \bibinfo {author} {\bibfnamefont
  {H.}~\bibnamefont {Kotegawa}}, \ and\ \bibinfo {author} {\bibfnamefont
  {J.}~\bibnamefont {Flouquet}},\ }\href@noop {} {\  (\bibinfo {year}
  {2010})},\ \Eprint {http://arxiv.org/abs/cond-mat/1012.1987}
  {arXiv:cond-mat/1012.1987} \BibitemShut {NoStop}%
\bibitem [{\citenamefont {L\'{e}vy}\ \emph {et~al.}(2009)\citenamefont
  {L\'{e}vy}, \citenamefont {Sheikin}, \citenamefont {Grenier}, \citenamefont
  {Marcenat},\ and\ \citenamefont {Huxley}}]{Levy09}%
  \BibitemOpen
  \bibfield  {author} {\bibinfo {author} {\bibfnamefont {F.}~\bibnamefont
  {L\'{e}vy}}, \bibinfo {author} {\bibfnamefont {I.}~\bibnamefont {Sheikin}},
  \bibinfo {author} {\bibfnamefont {B.}~\bibnamefont {Grenier}}, \bibinfo
  {author} {\bibfnamefont {C.}~\bibnamefont {Marcenat}}, \ and\ \bibinfo
  {author} {\bibfnamefont {A.}~\bibnamefont {Huxley}},\ }\href@noop {}
  {\bibfield  {journal} {\bibinfo  {journal} {J. Phys. Cond. Matt.}\ }\textbf
  {\bibinfo {volume} {21}},\ \bibinfo {pages} {164211} (\bibinfo {year}
  {2009})}\BibitemShut {NoStop}%
\bibitem [{\citenamefont {Aoki}\ \emph {et~al.}(2011)\citenamefont {Aoki},
  \citenamefont {Sheikin}, \citenamefont {Matsuda}, \citenamefont {Taufour},
  \citenamefont {Knebel},\ and\ \citenamefont {Flouquet}}]{Aoki10}%
  \BibitemOpen
  \bibfield  {author} {\bibinfo {author} {\bibfnamefont {D.}~\bibnamefont
  {Aoki}}, \bibinfo {author} {\bibfnamefont {I.}~\bibnamefont {Sheikin}},
  \bibinfo {author} {\bibfnamefont {T.~D.}\ \bibnamefont {Matsuda}}, \bibinfo
  {author} {\bibfnamefont {V.}~\bibnamefont {Taufour}}, \bibinfo {author}
  {\bibfnamefont {G.}~\bibnamefont {Knebel}}, \ and\ \bibinfo {author}
  {\bibfnamefont {J.}~\bibnamefont {Flouquet}},\ }\href@noop {} {\bibfield
  {journal} {\bibinfo  {journal} {J. Phys. Soc. Japan}\ }\textbf {\bibinfo
  {volume} {80}},\ \bibinfo {pages} {013705} (\bibinfo {year}
  {2011})}\BibitemShut {NoStop}%
\bibitem [{\citenamefont {Blanter}\ \emph {et~al.}(1994)\citenamefont
  {Blanter}, \citenamefont {Kaganov}, \citenamefont {Pantsulaya},\ and\
  \citenamefont {Varlamov}}]{Blanter94}%
  \BibitemOpen
  \bibfield  {author} {\bibinfo {author} {\bibfnamefont {Y.~M.}\ \bibnamefont
  {Blanter}}, \bibinfo {author} {\bibfnamefont {M.~I.}\ \bibnamefont
  {Kaganov}}, \bibinfo {author} {\bibfnamefont {A.~V.}\ \bibnamefont
  {Pantsulaya}}, \ and\ \bibinfo {author} {\bibfnamefont {A.~A.}\ \bibnamefont
  {Varlamov}},\ }\href@noop {} {\bibfield  {journal} {\bibinfo  {journal}
  {Physics Reports}\ }\textbf {\bibinfo {volume} {245}},\ \bibinfo {pages}
  {159} (\bibinfo {year} {1994})}\BibitemShut {NoStop}%
\bibitem [{\citenamefont {Katsnelson}\ and\ \citenamefont
  {Trefilov}(2000)}]{Katsnelson00}%
  \BibitemOpen
  \bibfield  {author} {\bibinfo {author} {\bibfnamefont {M.~I.}\ \bibnamefont
  {Katsnelson}}\ and\ \bibinfo {author} {\bibfnamefont {A.~V.}\ \bibnamefont
  {Trefilov}},\ }\href@noop {} {\bibfield  {journal} {\bibinfo  {journal}
  {Phys. Rev. B}\ }\textbf {\bibinfo {volume} {61}},\ \bibinfo {pages} {1643}
  (\bibinfo {year} {2000})}\BibitemShut {NoStop}%
\bibitem [{\citenamefont {Hackl}\ and\ \citenamefont {Vojta}(2011)}]{Hackl11}%
  \BibitemOpen
  \bibfield  {author} {\bibinfo {author} {\bibfnamefont {A.}~\bibnamefont
  {Hackl}}\ and\ \bibinfo {author} {\bibfnamefont {M.}~\bibnamefont {Vojta}},\
  }\href@noop {} {\bibfield  {journal} {\bibinfo  {journal} {Phys. Rev. Lett.}\
  }\textbf {\bibinfo {volume} {106}},\ \bibinfo {pages} {137002} (\bibinfo
  {year} {2011})}\BibitemShut {NoStop}%
\bibitem [{\citenamefont {Julian}\ \emph {et~al.}(1992)\citenamefont {Julian},
  \citenamefont {Teunissen},\ and\ \citenamefont {Wiegers}}]{Julian92}%
  \BibitemOpen
  \bibfield  {author} {\bibinfo {author} {\bibfnamefont {S.~R.}\ \bibnamefont
  {Julian}}, \bibinfo {author} {\bibfnamefont {P.~A.~A.}\ \bibnamefont
  {Teunissen}}, \ and\ \bibinfo {author} {\bibfnamefont {S.~A.~J.}\
  \bibnamefont {Wiegers}},\ }\href@noop {} {\bibfield  {journal} {\bibinfo
  {journal} {Phys. Rev. B}\ }\textbf {\bibinfo {volume} {46}},\ \bibinfo
  {pages} {9821} (\bibinfo {year} {1992})}\BibitemShut {NoStop}%
\bibitem [{\citenamefont {van Ruitenbeek}\ \emph {et~al.}(1982)\citenamefont
  {van Ruitenbeek}, \citenamefont {Verhoef}, \citenamefont {Mattocks},
  \citenamefont {Dixon}, \citenamefont {van Deursen},\ and\ \citenamefont
  {de~Vroomen}}]{vanRuitenbeek82}%
  \BibitemOpen
  \bibfield  {author} {\bibinfo {author} {\bibfnamefont {J.~M.}\ \bibnamefont
  {van Ruitenbeek}}, \bibinfo {author} {\bibfnamefont {W.~A.}\ \bibnamefont
  {Verhoef}}, \bibinfo {author} {\bibfnamefont {P.~G.}\ \bibnamefont
  {Mattocks}}, \bibinfo {author} {\bibfnamefont {A.~E.}\ \bibnamefont {Dixon}},
  \bibinfo {author} {\bibfnamefont {A.~P.~J.}\ \bibnamefont {van Deursen}}, \
  and\ \bibinfo {author} {\bibfnamefont {A.~R.}\ \bibnamefont {de~Vroomen}},\
  }\href@noop {} {\bibfield  {journal} {\bibinfo  {journal} {J. Phys. F: Met.
  Phys.}\ }\textbf {\bibinfo {volume} {12}},\ \bibinfo {pages} {2919} (\bibinfo
  {year} {1982})}\BibitemShut {NoStop}%
\bibitem [{\citenamefont {Mercure}\ \emph {et~al.}(2010)\citenamefont
  {Mercure}, \citenamefont {Rost}, \citenamefont {O'Farrell}, \citenamefont
  {Goh}, \citenamefont {Perry}, \citenamefont {Sutherland}, \citenamefont
  {Grigera}, \citenamefont {Borzi}, \citenamefont {Gegenwart}, \citenamefont
  {Gibbs},\ and\ \citenamefont {Mackenzie}}]{Mercure10}%
  \BibitemOpen
  \bibfield  {author} {\bibinfo {author} {\bibfnamefont {J.-F.}\ \bibnamefont
  {Mercure}}, \bibinfo {author} {\bibfnamefont {A.~W.}\ \bibnamefont {Rost}},
  \bibinfo {author} {\bibfnamefont {E.~C.~T.}\ \bibnamefont {O'Farrell}},
  \bibinfo {author} {\bibfnamefont {S.~K.}\ \bibnamefont {Goh}}, \bibinfo
  {author} {\bibfnamefont {R.~S.}\ \bibnamefont {Perry}}, \bibinfo {author}
  {\bibfnamefont {M.~L.}\ \bibnamefont {Sutherland}}, \bibinfo {author}
  {\bibfnamefont {S.~A.}\ \bibnamefont {Grigera}}, \bibinfo {author}
  {\bibfnamefont {R.~A.}\ \bibnamefont {Borzi}}, \bibinfo {author}
  {\bibfnamefont {P.}~\bibnamefont {Gegenwart}}, \bibinfo {author}
  {\bibfnamefont {A.~S.}\ \bibnamefont {Gibbs}}, \ and\ \bibinfo {author}
  {\bibfnamefont {A.~P.}\ \bibnamefont {Mackenzie}},\ }\href@noop {} {\bibfield
   {journal} {\bibinfo  {journal} {Phys. Rev. B}\ }\textbf {\bibinfo {volume}
  {81}},\ \bibinfo {pages} {235103} (\bibinfo {year} {2010})}\BibitemShut
  {NoStop}%
\bibitem [{\citenamefont {Rourke}\ \emph {et~al.}(2008)\citenamefont {Rourke},
  \citenamefont {McCollam}, \citenamefont {Lapertot}, \citenamefont {Knebel},
  \citenamefont {Flouquet},\ and\ \citenamefont {Julian}}]{Rourke08}%
  \BibitemOpen
  \bibfield  {author} {\bibinfo {author} {\bibfnamefont {P.~M.~C.}\
  \bibnamefont {Rourke}}, \bibinfo {author} {\bibfnamefont {A.}~\bibnamefont
  {McCollam}}, \bibinfo {author} {\bibfnamefont {G.}~\bibnamefont {Lapertot}},
  \bibinfo {author} {\bibfnamefont {G.}~\bibnamefont {Knebel}}, \bibinfo
  {author} {\bibfnamefont {J.}~\bibnamefont {Flouquet}}, \ and\ \bibinfo
  {author} {\bibfnamefont {S.~R.}\ \bibnamefont {Julian}},\ }\href@noop {}
  {\bibfield  {journal} {\bibinfo  {journal} {Phys. Rev. Lett.}\ }\textbf
  {\bibinfo {volume} {101}},\ \bibinfo {pages} {237205} (\bibinfo {year}
  {2008})}\BibitemShut {NoStop}%
\bibitem [{\citenamefont {Freeman}\ \emph {et~al.}(1976)\citenamefont
  {Freeman}, \citenamefont {Desclaux}, \citenamefont {Lander},\ and\
  \citenamefont {Faber}}]{Freeman76}%
  \BibitemOpen
  \bibfield  {author} {\bibinfo {author} {\bibfnamefont {A.~J.}\ \bibnamefont
  {Freeman}}, \bibinfo {author} {\bibfnamefont {J.~P.}\ \bibnamefont
  {Desclaux}}, \bibinfo {author} {\bibfnamefont {G.~H.}\ \bibnamefont
  {Lander}}, \ and\ \bibinfo {author} {\bibfnamefont {J.}~\bibnamefont
  {Faber}},\ }\href@noop {} {\bibfield  {journal} {\bibinfo  {journal} {Phys.
  Rev. B}\ }\textbf {\bibinfo {volume} {13}},\ \bibinfo {pages} {1168}
  (\bibinfo {year} {1976})}\BibitemShut {NoStop}%
\bibitem [{\citenamefont {Daou}\ \emph {et~al.}(2006)\citenamefont {Daou},
  \citenamefont {Bergemann},\ and\ \citenamefont {Julian}}]{Daou06}%
  \BibitemOpen
  \bibfield  {author} {\bibinfo {author} {\bibfnamefont {R.}~\bibnamefont
  {Daou}}, \bibinfo {author} {\bibfnamefont {C.}~\bibnamefont {Bergemann}}, \
  and\ \bibinfo {author} {\bibfnamefont {S.~R.}\ \bibnamefont {Julian}},\
  }\href@noop {} {\bibfield  {journal} {\bibinfo  {journal} {Phys. Rev. Lett.}\
  }\textbf {\bibinfo {volume} {96}},\ \bibinfo {pages} {026401} (\bibinfo
  {year} {2006})}\BibitemShut {NoStop}%
\bibitem [{\citenamefont {Kusminskiy}\ \emph {et~al.}(2008)\citenamefont
  {Kusminskiy}, \citenamefont {Beach}, \citenamefont {Neto},\ and\
  \citenamefont {Campbell}}]{Kusminskiy08}%
  \BibitemOpen
  \bibfield  {author} {\bibinfo {author} {\bibfnamefont {S.~V.}\ \bibnamefont
  {Kusminskiy}}, \bibinfo {author} {\bibfnamefont {K.~S.~D.}\ \bibnamefont
  {Beach}}, \bibinfo {author} {\bibfnamefont {A.~H.~C.}\ \bibnamefont {Neto}},
  \ and\ \bibinfo {author} {\bibfnamefont {D.~K.}\ \bibnamefont {Campbell}},\
  }\href@noop {} {\bibfield  {journal} {\bibinfo  {journal} {Phys. Rev. B}\
  }\textbf {\bibinfo {volume} {77}},\ \bibinfo {pages} {094419} (\bibinfo
  {year} {2008})}\BibitemShut {NoStop}%
\bibitem [{\citenamefont {Yamaji}\ \emph {et~al.}(2007)\citenamefont {Yamaji},
  \citenamefont {Misawa},\ and\ \citenamefont {Imada}}]{Yamaji07}%
  \BibitemOpen
  \bibfield  {author} {\bibinfo {author} {\bibfnamefont {Y.}~\bibnamefont
  {Yamaji}}, \bibinfo {author} {\bibfnamefont {T.}~\bibnamefont {Misawa}}, \
  and\ \bibinfo {author} {\bibfnamefont {M.}~\bibnamefont {Imada}},\
  }\href@noop {} {\bibfield  {journal} {\bibinfo  {journal} {J. Phys. Soc.
  Jpn.}\ }\textbf {\bibinfo {volume} {76}},\ \bibinfo {pages} {063702}
  (\bibinfo {year} {2007})}\BibitemShut {NoStop}%
\end{thebibliography}
\end{document}

\begin{center}
\textbf{Supplementary Information}
\end{center}

\textbf{Analysis of the quantum oscillation data.} The fast Fourier transforms in Fig.~2(b) show spectral weight below 200\,T in addition to the spectrally isolated peak that has formed the focus of the paper. At $\theta=6^\circ$ this weight clearly arises from the pronounced dip in the resistivity that is a remnant of the SC at lower angles. By 10$^\circ$ this dip has evolved into undulations that may signify another quantum oscillation component of very low frequency. The usual methods for distinguishing quantum oscillations from a background signal, namely studying the $T$, $B$ and angle dependence, are not helpful in this case because the background resistivity has a strong dependence on all these quantities that is unknown in detail and which arises both from enhanced scattering near the metamagnetic transition and from proximity to $B$-induced superconductivity.

To extract the quasiparticle mass and the field dependent quantum oscillation frequency of the clear quantum oscillation component at frequency 555\,T, we subtracted a smooth, slowly-varying background with negligible weight at the QO frequency. The oscillating residuals were fitted to a cosine function corresponding to Eq.~2, scaled so that its amplitude envelope described the data over the field range, and with the argument parameterized by a second order polynomial. This was done for a low $T$ trace with high signal-to-noise to give a definitive curve of oscillation phase versus $B$. The $B$-dependent frequency in Fig.~4(c) is calculated from this curve. Amplitudes as a function of $T$ and $B$ were found by fitting short $B$-windows of data, with the amplitude allowed to vary to optimize the fit. These amplitudes were fitted to $R_T$ to give the quasiparticle effective mass $m^*(B)$.

\textbf{Quantum oscillation amplitude on a shrinking Fermi surface pocket.} In general the amplitude of quantum oscillations can depend on intricate details of FS geometry but usually its $B$ dependence is a monotonically increasing function of magnetic field governed by the Dingle factor $R_\mathrm{D}=\exp(-\hbar\sqrt{\pi\mathcal{A}}/e B\ell_\mathrm{Q})$. We calculate the $B$-dependent amplitude of the resistance oscillations using Eqs. 1 and 2 incorporating the observed $B$ dependence of $m^*$ and $F$. This is done for two cases: (i) $\mathcal{A}(B)$ given by the middle curve in Fig.\ 4(d) corresponding to a nearly constant Fermi surface area; (ii) $\mathcal{A}(B)$ given by the lowest curve corresponding to a Fermi surface that vanishes via a Lifshitz transition at 15.5\,T. In Boltzmann transport theory the conductivity of the Fermi surface pocket can be written as $\sigma\propto S_\mathrm{F} v_\mathrm{F} \tau$ highlighting that a field dependent Fermi velocity and Fermi surface area $S_\mathrm{F}$ are both relevant. The temperatures of interest are low enough that $\tau$ is the $T\rightarrow 0$ limit of the transport lifetime and in both cases we take $\tau$ for the pocket to be independent of $B$. This can be consistent with dramatic changes in the size of the pocket if scattering on the pocket is predominantly inter-band and therefore determined by the DOS on other `spectator bands' with dominant DOS. For both cases we model the amplitude envelope of the oscillatory data by $\tilde{R}/|R|=C_0 \mathcal{A}(B)^{3/2} R_\mathrm{D}[\mathcal{A}(B),\ell_\mathrm{Q}] R_T[m^*(B)]/m^*(B)$ where $C_0$ is a free parameter controlling the overall amplitude. The factor of $\mathcal{A}^{3/2}/m^*$ comes from $S_\mathrm{F} v_\mathrm{F}$ and does not cancel in the ratio $\tilde{R}/|R|$ when spectator bands providing a parallel conduction channel are present. For case (i), the parameters $C_0$ and $\ell_\mathrm{Q}$ are chosen to fit the data up to $11$\,T but the evolution of the amplitude with magnetic field cannot be captured over the full field range by this model. In case (ii) the $B$-dependence of $\mathcal{A}^{3/2}$ makes is possible to choose $C_0$ and $\ell_\mathrm{Q}$ to reproduce the observed amplitude over the full range of magnetic field.

\textbf{Alternative explanations of non-monotonic amplitude.} We showed that a shrinking Fermi surface can explain the observed amplitude decrease at high fields. Other potential explanations can be eliminated as follows. Firstly a beat with another QO component of similar frequency would give  $B$ dependent modulations of the QO amplitude and not the abrupt vanishing observed at 15.5\,T. Other potential causes can be discussed with reference to the LK expression Eq.\ 2; the QO amplitude would decrease with magnetic field if (i) $R_\mathrm{D}$ is non-monotonic due to an increase in $\sqrt{F}/(\ell_\mathrm{Q}B)$ at high magnetic field, or (ii) the quasiparticle mass $m^*$ increases, or (iii) the local FS curvature, $\partial^2 \mathcal{A}/\partial k_\parallel^2$ increases. Case (i) implies a sudden decrease in $\ell_\mathrm{Q}$ beyond 15.5\,T, which is at odds with the decrease in resistivity. $F_\mathrm{obs}(B)$ is also not consistent with any $F(B)$ that would make $R_D$ maximum at $\sim 13\,T$. The observed decrease in $m^*$ eliminates (ii). Case (iii) would require large changes in the FS anisotropy; the $\theta$ independence of the observed QO frequency does not support this. A final possibility is that magnetic breakdown (MB) occurs at high $B$ and quasiparticles tunnel across an energy gap $\Delta$ to another section of FS provided $\hbar \omega_\mathrm{c}\gtrsim \Delta^2/\epsilon_\mathrm{F}$ where $\omega_\mathrm{c}$ is the cyclotron frequency \cite{ShoenbergBook}. MB can reduce QO amplitude as $B$ increases \cite{Dhillon55} but cannot explain the suddenness of the decrease observed.

\textbf{Calculation of $B_\mathrm{c2}^\mathrm{orb}$ and comparison with measured $B_\mathrm{c2}$.} To estimate $\xi$ for the $B_\mathrm{c2}^\mathrm{orb}$ calculation we assume the BCS expression $\xi=\hbar v_\mathrm{F}/\pi\Delta$ applies with $\Delta \approx 2 k_\mathrm{B} T_\mathrm{c}$ where $T_\mathrm{c}$ is the hypothetical zero-field superconducting critical temperature. The precise values of the numerical factors may differ from these since URhGe is likely to be a strongly coupled $p$-wave superconductor. In Fig.\ 5(b) the values of $B_\mathrm{c2}^\mathrm{orb}$ shown are for $B$ in the $b,c$-plane at $\theta=10^\circ$. The fact that $B_\mathrm{c2}^\mathrm{orb}$ is always less than the applied field is consistent with the absence of SC at $\theta=10^\circ$. In Fig.\ 5(c) we investigate whether the slowing down of quasiparticles approaching the Lifshitz transition can explain the emergence of the high-$B$ SC pocket without a divergence of the quasiparticle mass. To do this we fix the Lifshitz transition at the observed high-$B$ edge of the SC pocket and calculate $B_\mathrm{c2}^\mathrm{orb}$ for a field independent quasiparticle mass. The first result is that the values of the critical fields for $B\,||\,b$ within the model, namely the points where $B_\mathrm{c2}^\mathrm{orb}$ is exactly equal to the applied field, are in good agreement with observed values. To make a quantitative comparison of the model with experiment in the interval of $B_b$ where SC occurs for $B\,||\,b$, we calculate $B_\mathrm{c2}^\mathrm{orb}$ for field directions in the $a,b$-plane. For this we assume that for fields in the $a,b$-plane the Fermi surface size is tuned only by the component of $B$ parallel to $b$, which is consistent with the $a$-axis being magnetically very hard and the observation that $B_\mathrm{R}$ is independent of $B_a$ \cite{Levy07}. For a given value of $B_b$ the value of $B_\mathrm{c2}^\mathrm{orb}$ shown in the plot is the value for the field direction where it exactly equals the applied field, that is the model prediction for the critical field $B_\mathrm{c2}(B_b)$. The calculation assumes a $B$-independent coherence length anisotropy as deduced from the $B_\mathrm{c2}$ anisotropy of the low-field SC pocket of a lower quality sample that is expected to give the effective mass anisotropy encompassing both Fermi surface and gap anisotropy. The values used are $B_\mathrm{c2}^\mathrm{(LF,a)}=1.88$\,T, $B_\mathrm{c2}^\mathrm{(LF,b)}=1.22$\,T, $B_\mathrm{c2}^\mathrm{(LF,c)}=0.52$\,T. The angle dependent orbital limiting field is assumed to be of the standard form $B_\mathrm{c2}^\mathrm{orb}(\gamma,B_b)=B_\mathrm{c2}^\mathrm{orb}(0,B_b)/\sqrt{\cos^2 \gamma + \left(B_\mathrm{c2}^\mathrm{(LF,b)}/B_\mathrm{c2}^\mathrm{(LF,a)}\right)^2 \sin^2 \gamma}$ where $\gamma$ is the angle from $b$ within the $a,b$-plane.

%Discussion of crit field parallel to a
%Control: key (0)
%Control: author (72) initials jnrlst
%Control: editor formatted (1) identically to author
%Control: production of article title (-1) disabled
%Control: page (0) single
%Control: year (1) truncated
%Control: production of eprint (0) enabled
%